\documentclass[prl,aps,superscriptaddress,floats,twocolumn]{revtex4}
\usepackage{graphicx}
\usepackage{epsfig}
\begin{document}

\title{Searching for modifications to the exponential radioactive decay law 
with the Cassini spacecraft}       

\author{Peter.~S.~Cooper\\
        Fermi National Accelerator Laboratory, Batavia, IL 60510, U.S.A.}

\date{\today}

\begin{abstract}
Data from the power output of the radioisotope thermoelectric generators
aboard the Cassini spacecraft are used to test the conjecture that 
small deviations observed in terrestrial measurements of the exponential 
radioactive decay law are correlated with the Earth-Sun distance.  No 
significant deviations from exponential decay are observed over a range of 
$0.7 - 1.6 A.U.$ A $90\%$ Cl upper limit of $0.84\times10^{-4}$ is set on
a term in the decay rate of $^{238}Pu$ proportional to $1/R^{2}$ and
$0.99\times10^{-4}$ for a term proportional to $1/R$.

\end{abstract}

\maketitle

%\twocolumn

% Introduction
%
A recent archive preprint reports evidence for a correlation between nuclear 
decay rates and the Earth-Sun distance\cite{Fischbach}.   This correlation is
extracted from an annual modulation in the observed decay rates
of $^{32}Si / ^{36}Cl$, both $\beta$ emitters, and $^{226}Ra$, an
$\alpha$ emitter.  Reference~1 analyzes this as a correlation with 
$1/R^{2}(t)$, the variation in the Earth-Sun distance due to the eccentricity 
of the Earth's orbit.  To set a scale for this correlation the amplitude of
the decay rate variation is $\sim0.1\%$ roughly in phase with the 
$3\%$ annual modulation in $1/R^{2}(t)$ suggesting a $(3\times10^{-2})/R^{2}$ 
term in the decay rate.

In the conclusion of this paper the 
authors observe that: \emph{These conclusions can be tested...} [by] 
\emph{measurements on radioactive samples carried aboard spacecraft to other 
planets} [which] \emph{would be very useful since the sample-Sun 
distance would vary over a much wider range}.
I report here the results of exactly such a measurement based on the 
power output of the Radioisotope Thermoelectric Generators (RTG) aboard
the Cassini spacecraft which launched in 1997 and reached Saturn in 2004.

% Discussion
%

Cassini is powered by three RTGs each of which is a very large ($7.7Kg$, 
 $130KCu$) $^{238}Pu$ radioactive source, an $\alpha$ emitter with 
an $87.7y$ half life\cite{NNDC}.  The heat from these sources are converted to 
electric power with thermoelectric piles.  Together these sources produced 
$878w$ of electrical power from $\sim13Kw$ of radioactive decay heat at 
launch.  The power output of these RTGs were literally the lifeblood of the
Cassini mission.  Their power output was monitored carefully and often.

\begin{figure}
\includegraphics[width=0.5\textwidth]{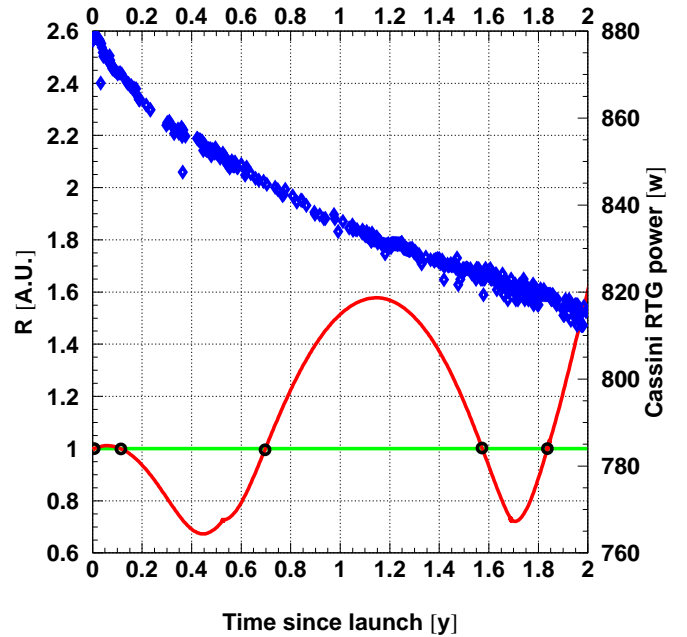}
\caption{Left, suppressed zero, scale: solid red curve, Heliocentric distance 
         [$R(t) (A.U.)$], Black points are the 5 times when R(t)=1 
         (green line),
         Right, suppressed zero, scale: blue diamonds, RTG electrical power.}
 \end{figure}

The trajectory of the Cassini spacecraft is available on the 
web\cite{TRAJ}.  I've used these data, converted to astronomical 
units ($A.U.$), to compute 
$1/R(t)^{2} = R_{e}^{2}/[x(t)^{2}+y(t)^{2}+z(t)^{2}]$.
Over the first 2 years Cassini went from R(0)=1 at launch, made 2 visits to 
Venus at R=0.7 and crossed the orbit of the earth a total of 4 times before
finally gaining enough speed to reach Saturn at R=9.  

JPL kindly provided\cite{JPL} P(t), the total electrical power from 
the three RTGs aboard, measured daily since launch and the expected 
power output from their RTG modeling.  The distance of Cassini from the Sun 
and the electrical power output are plotted in Figure~1 for the first $2$ 
years of the mission.  R ranged from $0.7 - 1.6 A.U.$ $(0.35<1/R^{2}<2.2)$.
The power dropped from $878w$ at launch to $815w$ over this period.

The thermal power output of an RTG is directly proportional to the decay rate
of the radioisotope generating the heat: 
$P_{th}(t) = N_{0}\lambda exp(-\lambda t) E_{d}$, where $E_{d}$ 
is the energy released per decay. The electrical power output, 
$P(t) = P_{th}(t) \epsilon_{0} \epsilon(t)$ 
is modified by an initial efficiency, $\epsilon_{0}$, and a time (or power) 
dependent thermoelectric conversion efficiency; $\epsilon(t), \epsilon(0)=1$.
  
I cannot safely use the RTG efficiency model\cite{DEGRA} here since it assumed 
only exponential behavior for radioactive decay.  Any new physics effect 
might be inadvertently subsumed into that model.  The Cassini trajectory 
provides a natural calibration for $\epsilon(t)$ using the measured power at 
the 5 points (shown in Figure~1) where the spacecraft was $1 A.U.$ from the 
sun.  Fitting these measurements to 
$P_{R=1}(t) = P(0) 2^{(-t/87.7y)} 2^{(-t/T_{eff})}$ yields 
$T_{eff}=21.2\pm1.9y$.  This simple model agrees in shape with 
Reference~4 and obeys the requirements of Carnot thermal efficiency: as the 
power decreases, and the temperature difference across the thermoelectric piles
decrease, the efficiency can only decrease. Something as complicated as a 
space-born thermoelectric pile requires more than one parameter to accurately 
describe its behavior.  The single exponential reduces a $5\%$ power drop in 
the first 2 years to a $\pm1\%$ variation from unity.

Following Reference~1 I've plotted 
$\epsilon(t)U(t) = P(t)/P(0) 2^{ t/87.7y}$, the normalized electrical power 
corrected for $^{238}Pu$ decay, and U(t), the normalized thermal power, as a 
functions of time since launch in Figure~2 for the first 2 years of Cassini's 
voyage to Saturn.

\begin{figure}
\includegraphics[width=0.5\textwidth]{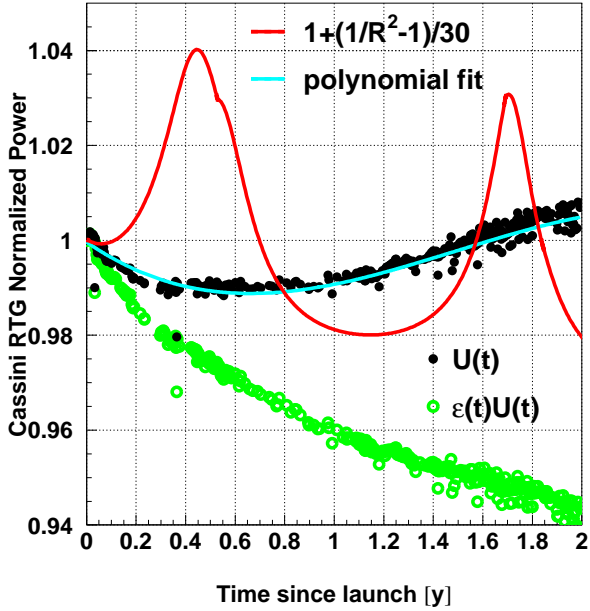}
\caption{ RTG power;
            green open circles, decay corrected [$\epsilon(t)U(t)$],
            black solid circles, efficiency and decay corrected [$U(t)$],
            cyan curve, 3rd order polynomial correction function described 
            the in text.
            red solid line, expected effect extrapolated from Reference~1:
            $1-(1/R^{2}(t)-1)/30$.}
 
\end{figure}

To compare to the rough magnitude of the effect reported by 
reference 1 I've plotted $U_{ref 1}(t) = 1+(0.1\%/3\%)(1/R(t)^{2}-1)$ on 
Figure~2 to extrapolate the small R variation of reference 1 for comparison 
with the larger R range available in this study.  Extrapolating a 
$0.1\%$ decay rate change for a $3\%$ change in $1/R^{2}$ to a $50\%$ change 
in $1/R^{2}$ ($R=0.7A.U.$, $t=0.43y$) should cause a $+4\%$ change in 
the power output of the RTG 5 months after launch.  In fact $U(t)$
decreases by $\sim1\%$.  Changes this large are excluded by the 
Cassini efficiency corrected data both in magnitude and by the absence of any 
reflection of the shape of $1/R^{2}(t)$ in either normalized power curve.  

% Limits
%
In order to set quantitative limits I've fit the efficiency and half-life
corrected $U(t)$ normalized power data (suppressing the two obvious outlying
points in Figure 1) to $P_{3}(t)$, a 3th order polynomial 
in time, to phenomenologically describe the last $1\%$  variation in the 
$U(t)$.   As shown in Figure~2, this polynomial smoothly interpolates the 
$U(t)$ measurements.  A 3th order polynomial make a very poor fit to 
$1/R^2(t)$; these two shapes are approximately orthogonal.  The error assigned 
to each U(t) measurement by requiring this fit to have $\chi^{2}/\nu = 1$ is 
$0.0015$.  The individual relative power measurements have a resolution of 
$0.15\%$ and the polynomial is a $\sim6 \sigma$  systematic correction beyond 
the simple exponential efficiency model.

\begin{figure}
\includegraphics[width=0.5\textwidth]{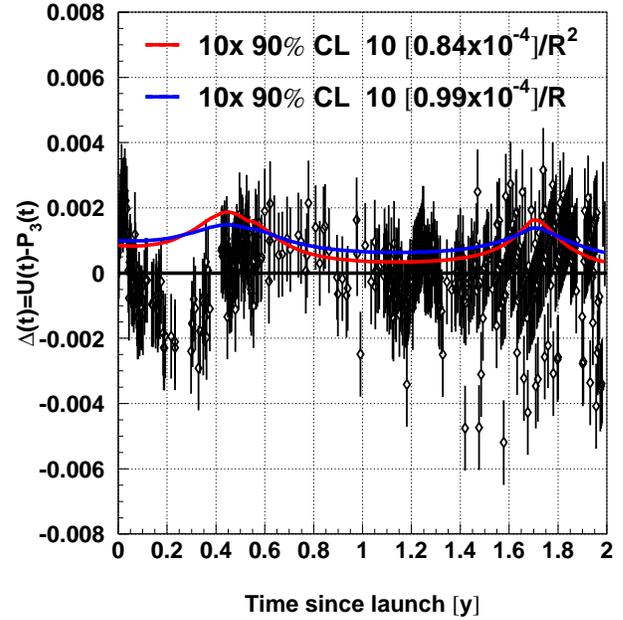}
\caption{Black points; $\Delta(t)$, RTG normalized thermal power less a 
            3rd order polynomial correction, 
            Red curve;  $10$ times the $90\%$ CL limit, $10\alpha/R^{2}$, 
            Blue curve; $10$ times the $90\%$ CL limit, $10\beta/R$.}
\end{figure}

The difference of the data from the polynomial fit; 
$\Delta(t) = U(t) - P_{3}(t)$ are plotted in Figure~3.  Some structure at the
$1 \sigma$ level and some outlying measurements remain.  These difference data 
are fit to $\alpha/R^{2}$ and $\beta/R$ to give limits on the contribution of 
a term in the $^{238}Pu$ decay rate dependent on the Earth-Sun distance.  
The $90\% CL$ limit on from these fits are $|\alpha|<0.84\times10^{-4}$ and 
$|\beta|<0.99\times10^{-4}$ respectively.   The limiting functions, scale up 
by a factor of $10$ for visibility, are also shown in Figure 3.  $\alpha$ is 
to be compared with the correlation seen in  Reference~1 for $^{226}Rn$ decay 
of $\sim+3\times10^{-2}$.  

% Conclusions
%
The Cassini RTG power data exclude any variation of the $^{238}Pu$ nuclear 
decay rate correlated with the distance of the source from the Sun to a 
level $350\times$ smaller than the effect reported by Reference 1.  $^{238}Pu$
and $^{226}Ra$ are similar $\alpha$ emitters.  Another physical or 
experimental cause of the reported annual variations in nuclear decay rates 
appears to be necessary.  More generally Rutherford, Chadwick, and Ellis's 
1930 conclusion that \emph{The rate of transformation of an element has been 
found to be constant under all conditions.}\cite{Rutherford} now has solid 
experimental support at least from Venus (R=0.7) to Mars (R=1.5).

% Acknowledgement_paragraph.tex
%
I am indebted to several of my colleagues for calling this
paper and physics issue to my attention; Chris Quigg and Martin Hu of Fermilab
and Jurgen~Engelfried of the Universidad Aut\'{\o}noma de San Luis 
Potos\'{\i}, Mexico.  I am also thankful for very useful 
lunchtime conversations on this subject with several of my Fermilab 
colleagues.  I an indebted to Richard Ewell and Torrence Johnson of JPL for
making the RTG data available and Stephen Parke of Fermilab for critical 
comments on this manuscript. 


\begin{thebibliography}{9}

\bibitem{Fischbach} Jere~H.~Jenkins {\sl et al.},
                 \emph{Evidence for Correlations Between Nuclear Decay 
                    Rates and the Earth-Sun Distance},
                 arXiv:astro-ph:0808.3283v1. 
                 % submitted to Phys.Rev.Lett.


\bibitem{NNDC}   National Nuclear Data Center, http://www.nndc.bnl.gov/.

\bibitem{TRAJ}   http://www.lepp.cornell.edu/~seb/celestia/cassini-all.zip

\bibitem{JPL}    private communication, R.~Ewell, JPL. 

\bibitem{DEGRA}  R.~Ewell, D.~Hanks, J.~Lozano, V.~Shields, 
                 $\&$ E.~Wood,
                 \emph{DEGRA - A computer Model for Predicting Long 
                       Term Thermoelectric Generator Performance},
                  Space Technologies and Applications International 
                  Forum (STAIF), Albuquerque, New Mexico, 
                  February 12-16, 2005., 
                  [http://hdl.handle.net/2014/38760]

\bibitem{Rutherford}    S.~E.~Rutherford, J.~Chadwick, and C.~Ellis, 
                  \emph{Radiations from Radioactive Substances} 
                  (Cambridge University Press, 1930).


\end{thebibliography}
\end{document}